# Quantum dynamics of wave packets in a nonstationary parabolic potential and the Kramers escape rate theory

Vladimir I. Dubinko[1*], Denis V. Laptev[2], Alexander S. Mazmanishvili[1], Juan F. R. Archilla[3]

[1] *National Science Center "Kharkov Institute of Physics and Technology", Kharkov 61108, Ukraine*
[*]vdubinko@hotmail.com
[2] *B. Verkin Institute for Low Temperature Physics and Engineering, Kharkov 61103, Ukraine*
[3] Group of Nonlinear Physics, Universidad de Sevilla, Sevilla 41011, Spain

At sufficiently low temperatures, the reaction rates in solids are controlled *by quantum* rather than by thermal fluctuations. We solve the Schrödinger equation for a Gaussian wave packet in a *nonstationary* harmonic oscillator and derive simple analytical expressions for the increase of its mean energy with time induced by the time-periodic modulation. Applying these expressions to the modified Kramers theory, we demonstrate a strong increase of the rate of escape out of a potential well under the time-periodic driving, when the driving frequency of the well position equals its eigenfrequency, or when the driving frequency of the well width exceeds its eigenfrequency by a factor of ~2. Such regimes can be realized near localized anharmonic vibrations (LAVs), in which the amplitude of atomic oscillations greatly exceeds that of harmonic oscillations (phonons) that determine the system temperature. LAVs can be excited either thermally or by external triggering, which can result in strong catalytic effects due to amplification of the Kramers rate.

## 1. Introduction

Theory of escape rates over a potential barrier, first proposed by Kramers in 1940 [1] is an archetype model for chemical reactions, and it has many applications in chemistry kinetics, diffusion in solids, nucleation and other phenomena [2]. The model considers a Brownian particle moving in a symmetric double-well potential $U(x)$ (Fig. 1a). The position of the particle represents the (free) energy of a system including the 'reaction site' in the phase space *energy-reaction coordinate*. The particle is subject to fluctuational forces that cause transitions between the neighboring potential wells with a rate given by the celebrated Kramers rate:

$$R_K = \frac{\omega_0}{2\pi} \exp\left[-E_0/D(T)\right], \quad \omega_0^2 = U''(x_m)/m \tag{1}$$

where $\omega_0/2\pi$ is the natural attempt frequency, $\omega_0$ being the angular frequency of the harmonic oscillator determined by the curvature of the first energy minimum (*i.e.* the reactants), and $E_0$ is the height of the potential barrier separating the two stable states, corresponding to the reactants and products,

$D(T)$ is the strength of the Gaussian white noise induced by thermal and quantum fluctuations. In a general case, $D(T)$ is given by [3]

$$D(T) = E_{ZPO} \coth(E_{ZPO}/k_B T) \approx \begin{cases} E_{ZPO}, & T \to 0 \\ k_B T, & T \gg E_{ZPO}/k_B \end{cases}, \quad E_{ZPO} = \frac{\hbar \omega_0}{2}, \qquad (2)$$

where $E_{ZPO}$ is the energy of *zero-point oscillations*, i.e. the ground state energy of the harmonic oscillator, $\hbar$ is the Plank constant, $k_b$ is the Boltzmann constant and $T$ is the temperature. It is quite natural to use the noise strength (2) in the calculation of the Kramers escape rate of the potential well, which results in a non-vanishing escape rate even if $T \to 0$ [3]. At sufficiently high temperatures, the noise strength becomes equal to $k_B T$, and the Kramers rate (1) provides a theoretical basis for the Arrhenius law, whereas at low temperatures, significant deviations from this low are predicted due to quantum effects.

The original Kramers model assumes a stationary potential landscape for a Brownian particle, which can be questioned in the situations where the reaction site is in the vicinity of *localized anharmonic vibrations* (LAVs) of atoms known also as 'discrete breathers' [4-8] or 'intrinsic localized modes' [9, 10] arising in regular crystals. In contrast to phonons, LAVs are large amplitude and *periodic in time*, and therefore they can induce a time-periodic modulation (driving) of the reaction potential landscape. It can involve the time-periodic oscillation of the potential barrier *height* and *shape*. The former case was analyzed by Dubinko et al [4, 5] who showed that if the driving frequency $\Omega$ is about or lower than the natural frequency, $\Omega \leq \omega_0$, one can use an 'adiabatic' approximation. In this case, the reaction rate $\langle R_K \rangle$, averaged over times exceeding the driving period, has been shown to increase with respect to the ground value $R_K$ according to the following expression

$$\frac{\langle R_K \rangle}{R_K} \approx \frac{\Omega}{2\pi} \int_0^{2\pi/\Omega} \exp\left(\frac{V \cos(\Omega t)}{k_b T}\right) dt = I_0\left(\frac{V}{k_b T}\right), \qquad (3)$$

where the amplification factor can be approximated by the zero order, modified Bessel function of the first kind $I_0(x)$ with the argument determined by the ratio of the driving amplitude $V$ to the temperature, and it weakly depends on the driving frequency or the barrier height [5].

Figure 1b shows that the LAV-induced periodic driving of the barrier height can amplify the average reaction rate drastically if the ratio $V/k_B T$ is high enough. That is expected to be the case in the reaction site interacting with a nearby LAV, since MD simulations using realistic interatomic potentials of various materials show that a typical deviation of the potential energy of atoms within a LAV is of the order of several fractions of eV [7-10].

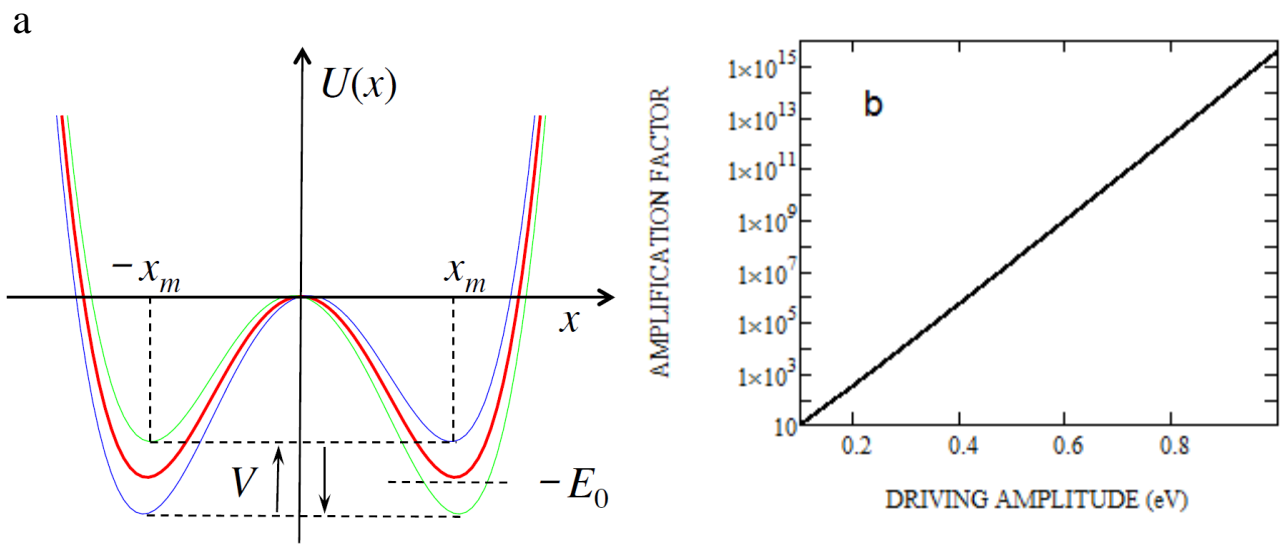

**Fig. 1.** (a) Sketch of the double-well potential $U(x) = (1/4)bx^4 - (1/2)ax^2$ (red curve) The minima are located at $\pm x_m$, where $x_m = (a/b)^{1/2}$. These are stable states before and after reaction, separated by a potential 'barrier' with the height $E_0 = a^2/4b$ changing periodically within the $V$ band. The green and blue curves represent the two maximally tilted energy landscapes. (b) Amplification factor for the escape rate of a thermalized Brownian particle at $T$ = 300 K from a periodically driven potential well vs. the driving amplitude, according to eq. (3) [5].

In the present paper, we consider effects due to the periodic driving of the potential landscape *shape*, which are not taken into account by the amplification factor (3) that is sensitive only to the modulation of the barrier *height*. Driving of the potential landscape shape (without changing the barrier height), can result in the time-periodic modulation (i) of the curvature and (ii) of the positions of the potential minima. We will consider the corresponding effects separately, since the quantum dynamics of the oscillating wave functions in these limiting cases are qualitatively different.

**2. Solution of the Schrödinger equation for a harmonic oscillator with time-dependent frequency**

Initial state of the system (reactants) can be described by a wave function of the Gaussian form placed near the first energy minimum that can be approximated by a parabolic potential [11]:

$$\psi(x_0, t_0 = 0) = \frac{1}{\sqrt[4]{\pi\xi^2}} \exp\left(-\frac{x_0^2}{2\xi^2}\right), \quad \xi = \sqrt{\hbar/m\omega_0}, \tag{4}$$

Time-periodic modulation of the potential curvature will result in time-periodic modulation of the harmonic oscillator eigenfrequency. A harmonic oscillator with time-dependent frequency for a particle with the mass *m* obeys the nonstationary Schrödinger equation of the form

$$i\hbar\frac{\partial \psi}{\partial t} = -\frac{\hbar^2}{2m}\frac{\partial^2 \psi}{\partial x^2} + \frac{m\omega^2(t)}{2}x^2\psi. \tag{5}$$

The solution of the equation (5) can be expressed using the Green`s function (or propagator) and the following initial condition [12]:

$$\psi(x,t) = \int_{-\infty}^{+\infty} dx_0 G(x,t;x_0,t_0)\psi(x_0,t_0), \quad \lim_{t \to \tau+0} G(x,t;x_0,t_0) = \delta(x - x_0) \tag{6}$$

The expression for the propagator has the form [12]:

$$G(x,t;x_0,t_0) = \sqrt{\frac{m}{2\pi i\hbar Z}}\exp(\theta_G), \quad \theta_G = \frac{im}{2\hbar Z}\left[\frac{dZ}{dt}x^2 - 2xx_0 + Yx_0^2\right], \tag{7}$$

where the functions $Y = Y(t)$, $Z = Z(t)$ are defined by the following equations and initial conditions that can be derived from the condition (7):

$$\frac{d^2Y}{dt^2} + \omega^2(t)Y = 0, \quad \frac{dY(t_0)}{dt} = 0, \quad Y(t_0) = 1, \tag{8}$$

$$\frac{d^2Z}{dt^2} + \omega^2(t)Z = 0, \quad \frac{dZ(t_0)}{dt} = 1, \quad Z(t_0) = 0, \tag{9}$$

$$Y(t)\frac{dZ(t)}{dt} - Z(t)\frac{dY(t)}{dt} = 1. \tag{10}$$

Then the expression for the wave function for the arbitrary moment of time $\forall t > t_0 = 0$ can be obtained in the following form:

$$\psi(x,t) = \frac{1}{\sqrt[4]{\pi\xi^2}}\frac{\exp(\theta_\psi)}{\sqrt{Y + i\omega_0 Z}}, \quad \theta_\psi = -\frac{x^2}{2\xi^2}\frac{1}{i\omega_0 Z}\left[\frac{dZ}{dt} - \frac{1}{Y + i\omega_0 Z}\right] \tag{11}$$

The probability density of finding the particle at *(x, t)* is given by the square of the wave function:

$$|\psi(x,t)|^2 = \frac{B(t)}{\xi\sqrt{\pi}} \exp\left\{-\frac{x^2}{\xi^2} B^2(t)\right\}, \quad B(t) = \frac{1}{\sqrt{Y^2(t) + \omega_0^2 Z^2(t)}}, \tag{12}$$

while dispersions of coordinate and momentum are given by:

$$\sigma_x(t) = \langle (x - \langle x \rangle)^2 \rangle = \frac{\hbar}{2m\omega_0}\left[Y^2 + \omega_0^2 Z^2\right], \tag{13}$$

$$\sigma_p(t) = \langle (p - \langle p \rangle)^2 \rangle = \frac{\hbar m\omega_0}{2}\left[\left(\frac{1}{\omega_0}\frac{dY}{dt}\right)^2 + \left(\frac{dZ}{dt}\right)^2\right], \tag{14}$$

At a *constant eigenfrequency:* $\omega(t) = \omega_0 = const$, the x and p dispersions are constant as well as the mean oscillator energy $\langle E \rangle = \frac{1}{2m}\sigma_p + \frac{m\omega^2}{2}\sigma_x$, i.e. the zero point oscillator energy $E_{ZPO}$ and the maximum mean square displacement from the equilibrium position, i.e. the ZPO amplitude $A_{ZPO}$:

$$\sigma_x = \frac{\hbar}{2m\omega_0}, \quad \sigma_p = \frac{\hbar m\omega_0}{2}, \quad \langle E \rangle = E_{ZPO} = \frac{\hbar\omega_0}{2}, \quad A_{ZPO} = \sqrt{\frac{\hbar}{2m\omega_0}}. \tag{15}$$

In a special case of *parametric time-periodic modulation* of the eigenfrequency with the driving frequency $\Omega = 2\omega_0$ considered in ref. [12], the equations (8), (9) are the Mathieu equations:

$$\ddot{x} + \omega_0^2\left[1 - g\cos(2\omega_0 t)\right]x = 0 \tag{16}$$

which solution can be written explicitly in the first approximation to the <u>small modulation amplitude</u> $g \ll 1$, and one obtains the first approximations for dispersion of the coordinate and momentum, which describe fully the evolution of the Gaussian wave packet in time:

$$\sigma_x(t) = \frac{\hbar}{2m\omega_0}\cosh\left(\frac{g\omega_0 t}{2}\right)\left[1 + \tanh\left(\frac{g\omega_0 t}{2}\right)\sin(2\omega_0 t)\right], \tag{17}$$

$$\sigma_p(t) = \frac{\hbar m\omega_0}{2}\cosh\left(\frac{g\omega_0 t}{2}\right)\left[1 - \tanh\left(\frac{g\omega_0 t}{2}\right)\sin(2\omega_0 t)\right], \tag{18}$$

In particular, the ZPO mean energy and amplitude increase with time as:

$$E_{ZPO}(t) = \frac{\hbar\omega_0}{2}\cosh\frac{g\omega_0 t}{2}, \quad A_{ZPO}(t) = \sqrt{\frac{\hbar}{2m\omega_0}\cosh\frac{g\omega_0 t}{2}}. \tag{19}$$

Fig. 2a shows that the parametric modulation of a parabolic potential well increases the dispersion of the wave packet with increasing number of oscillation periods, $N = \omega_0 t / 2\pi$, which results in rapidly increasing probability to find the oscillating particle far beyond the characteristic length of the stationary well $\xi = A_{ZPO}(0)\sqrt{2}$. Another significant effect of the modulation is the continuous increase of the ZPO energy (Fig. 2b), which is different from the *quantum* energy increase to the higher oscillation levels: $E_n = \hbar\omega_0 (1/2 + n)$, when the probability density becomes concentrated at the classical "turning points". In contrast to that, we clearly deal with the ground (zero-point) state, in which the probability density is concentrated at the origin, which means that the particle spends most of its time at the bottom of the potential well, whereas the dispersion of its position and momentum increases along with its zero-point energy due to the parametric modulation.

Although analytical solution of the problem for an arbitrary modulation frequency cannot be obtained, numerical analysis shows that the maximum effect is produced by a parametric modulation at $\Omega = 2\omega_0$. In the following section, we will consider another type of the harmonic well modulation that conserves its eigenfrequency but changes the position of the well minimum.

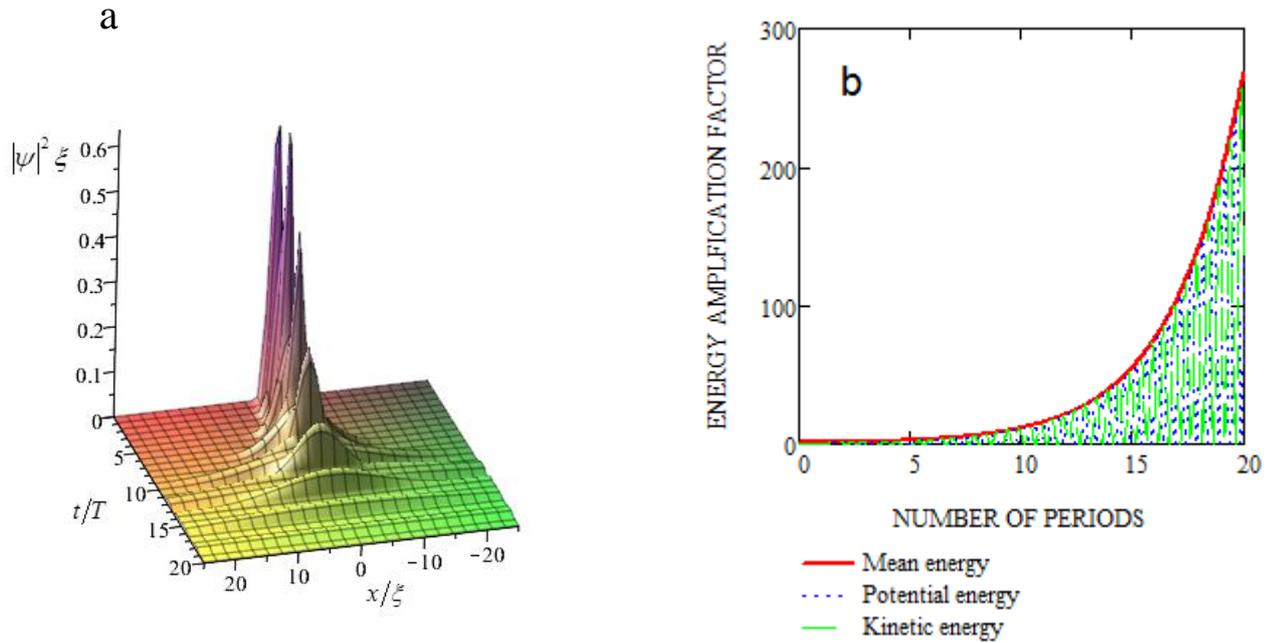

**Fig. 2**. (a) Localization probability distribution vs. the number of oscillation periods $N = \omega_0 t / 2\pi = t/T$ in the parametric regime $\Omega = 2\omega_0$ at $g = 0.1$ according to eq. (12). (b) Ratio of the zero-point energy to its stationary value in the parametric regime at $g = 0.1$ according to eq. (19).

## 3. Solution of the Schrödinger equation for a harmonic oscillator with time-dependent position of the potential minimum

In this case, the Schrödinger equation takes the following form:

$$i\hbar \frac{\partial}{\partial t}\psi(x,t) = -\frac{\hbar^2}{2m}\frac{\partial^2}{\partial x^2}\psi(x,t) + \frac{m\omega_0^2}{2}\left[x - X(t)\right]^2 \psi(x,t). \tag{20}$$

Its solution is given by eq. (6) with the Green function of the form [13]

$$G(x,t;x_0,t_0=0) = \sqrt{\frac{m\omega_0}{2\pi i\hbar \sin(\omega_0 t)}} \exp\{\theta(x,t;x_0,0)\}, \tag{21}$$

$$\theta(x,t;x_0,0) = -\frac{im\omega_0^2}{2\hbar}\int_0^t d\tau X^2(\tau) + \frac{i\hbar}{2m}\int_0^t d\tau a^2(\tau) + a(t)x +$$

$$+ \frac{m\omega_0}{\hbar}\frac{\left[x + s(t)\exp(i\omega_0 t)\right]^2}{1 - \exp(i2\omega_0 t)} - \frac{m\omega_0}{2\hbar}\left(x^2 - x_0^2\right), \tag{22}$$

$$a(t) = \frac{im\omega_0^2}{\hbar}\int_0^t d\tau X(\tau)\exp[i\omega_0(\tau - t)], \quad s(t) = -x_0 + \frac{i\hbar}{m}\int_0^t d\tau a(\tau)\exp(-i\omega_0 \tau), \tag{23}$$

Then for the initial wave packet given by eq. (4), an expression for the wave function for the arbitrary moment of time $\forall t > t_0 = 0$ can be obtained in the following form:

$$\psi(x,t) = i\sqrt[4]{\frac{m\omega_0}{\pi\hbar}} \exp\left\{-\frac{m\omega_0}{2\hbar}\left[x^2 + 2x\omega_0 \int_0^t d\tau X(\tau)\sin\omega_0(\tau - t) - F(t)\right] - iJ(x,t)\right\}, \tag{24}$$

$$F(t) = 2\omega_0^3 \int_0^t d\tau \left(\int_0^\tau d\tau' X(\tau')\sin\omega_0(\tau' - \tau)\right)\cdot\left(\int_0^\tau d\tau'' X(\tau'')\cos\omega_0(\tau'' - \tau)\right), \tag{25}$$

$$J(x,t) = \frac{m\omega_0^2}{2\hbar}\left\{\int_0^t d\tau U^2(\tau) + \frac{\hbar t}{m\omega_0} - \omega_0^2 I_1(t) - 2xI_2(t)\right\}, \tag{26}$$

$$I_1(t) = \int_0^t d\tau \left[\left(\int_0^\tau d\tau' U(\tau')\sin\omega_0(\tau' - \tau)\right)^2 - \left(\int_0^\tau d\tau' U(\tau')\cos\omega_0(\tau' - \tau)\right)^2\right], \tag{27}$$

$$I_2(t) = \int_0^t d\tau U(\tau)\cos\omega_0(\tau - t), \tag{28}$$

Accordingly, the probability density distribution is given by

$$|\psi(x,t)|^2 = \sqrt{\frac{m\omega_0}{\pi\hbar}} \exp\left\{-\frac{m\omega_0}{\hbar}\left[x^2 + 2x\omega_0\int_0^t d\tau X(\tau)\sin\omega_0(\tau-t) - F(t)\right]\right\}, \quad (29)$$

Time-periodic modulation has a maximum effect on the oscillator when the modulation frequency equals the eigenfrequency, when one has a wave packet concentrated around a 'centre of mass' with a coordinate $\lambda(t)$ that oscillates with amplitude linearly increasing in time (Fig.3 a) and the mean energy that increases as $t^2$ (Fig. 3b):

$$X(t) = g_A A_{ZPO}\sin(\omega_0 t), \quad \psi(x,t) = i\sqrt[4]{\frac{m\omega_0}{\pi\hbar}}\exp\left\{-\frac{m\omega_0}{2\hbar}[x+\lambda(t)]^2 - iJ(x,t)\right\}, \quad (30)$$

$$\lambda(t) = \frac{g_A A_{ZPO}}{2}\omega_0 t\left(\cos\omega_0 t - \frac{\sin\omega_0 t}{\omega_0 t}\right), \quad g_A \text{ - the modulation factor}, \quad (31)$$

$$\langle E\rangle = \frac{\hbar\omega_0}{2} + \frac{(g_A A_{ZPO})^2 m\omega_0^2}{8}\left[\omega_0^2 t^2 + \omega_0 t\sin 2\omega_0 t + \sin^2\omega_0 t\right], \quad A_{ZPO} = \sqrt{\frac{\hbar}{2m\omega_0}}, \quad (32)$$

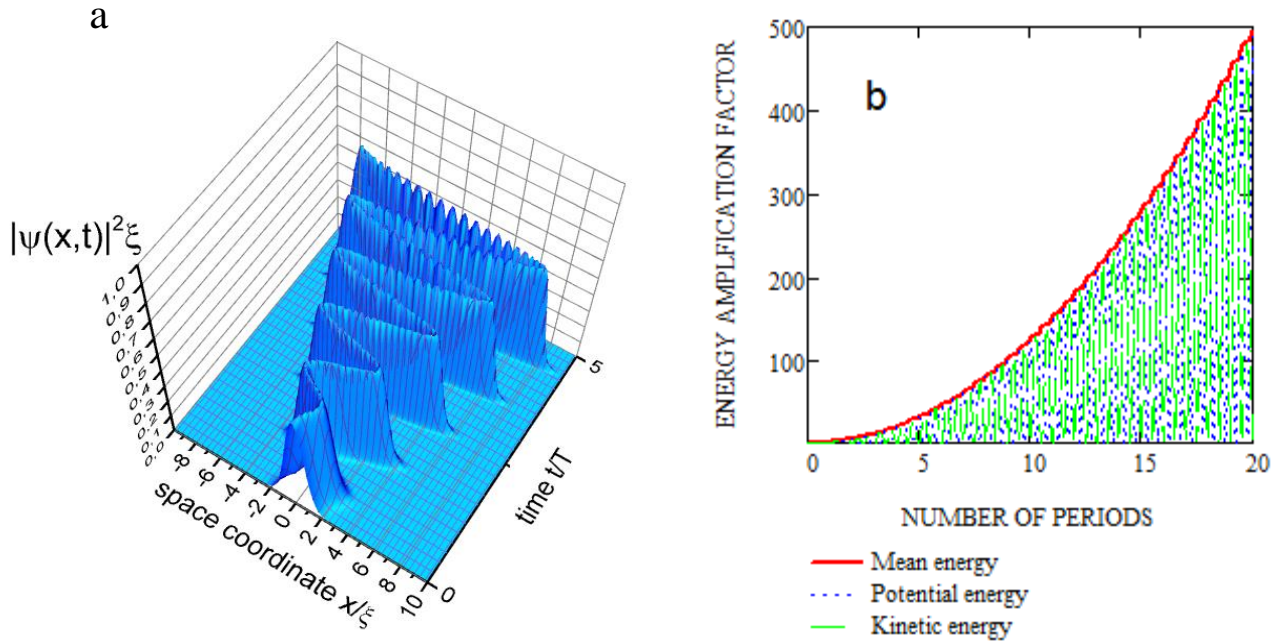

**Fig. 3**. (a) Localization probability distribution by eq. (31) and (b) ratio of the mean energy to its stationary value by eq. (32) vs. the number of oscillation periods $N = \omega_0 t/2\pi = t/T$ at $\Omega = \omega_0$, $g_A = 0.5$.

The driving of the well position does not affect the wave packet dispersions in the p and x space that remain constant:

$$\sigma_x = \frac{\hbar}{2m\omega_0}, \qquad \sigma_p = \frac{\hbar m\omega_0}{2}, \qquad A_{ZPO} = \sqrt{\frac{\hbar}{2m\omega_0}}, \tag{33}$$

which means that the wave packet deviates from the well bottom as a whole, and the uncertainty of the particle position does not increase with time, in a marked contrast to the well eigenfrequency modulation considered in the previous section (Fig.2.).

In the following section, we will discuss the applications of the above analysis of quantum dynamics of non-stationary harmonic oscillators to the modified Kramers rate of escape out of a potential well with account of the time-periodic driving.

**4. Discussion**

It is known that tunnel effect is inherently related to the operation of the uncertainty principle similar to the ZPO energy, the difference being that for the tunnel effect the coordinate is one in which the potential energy passes through a maximum, whereas for ZPO energy it passes through a minimum [14]. This thesis is well illustrated by the Kramers model of escape out of a potential well modified with account of ZPO energy [3], according to which the strength of the Gaussian white noise is determined by a synergetic action of *thermal* and *quantum* fluctuations, as described by eq. (2).

In Section 2, we have demonstrated that the parametric modulation of the well eigenfrequency increases the *strength of quantum fluctuations* manifested by the wave packet broadening and by the increase of its ZPO energy given by eq. (19). Substituting eq. (19) into eqs. (1), (2) one can evaluate the rate of escape from a well with a given parameters, which will increase with increasing number of oscillation periods, as illustrated in Fig. 4a for a potential well with a depth of 1 eV, and with eigenfrequency ranging from 5 to 50 THz.

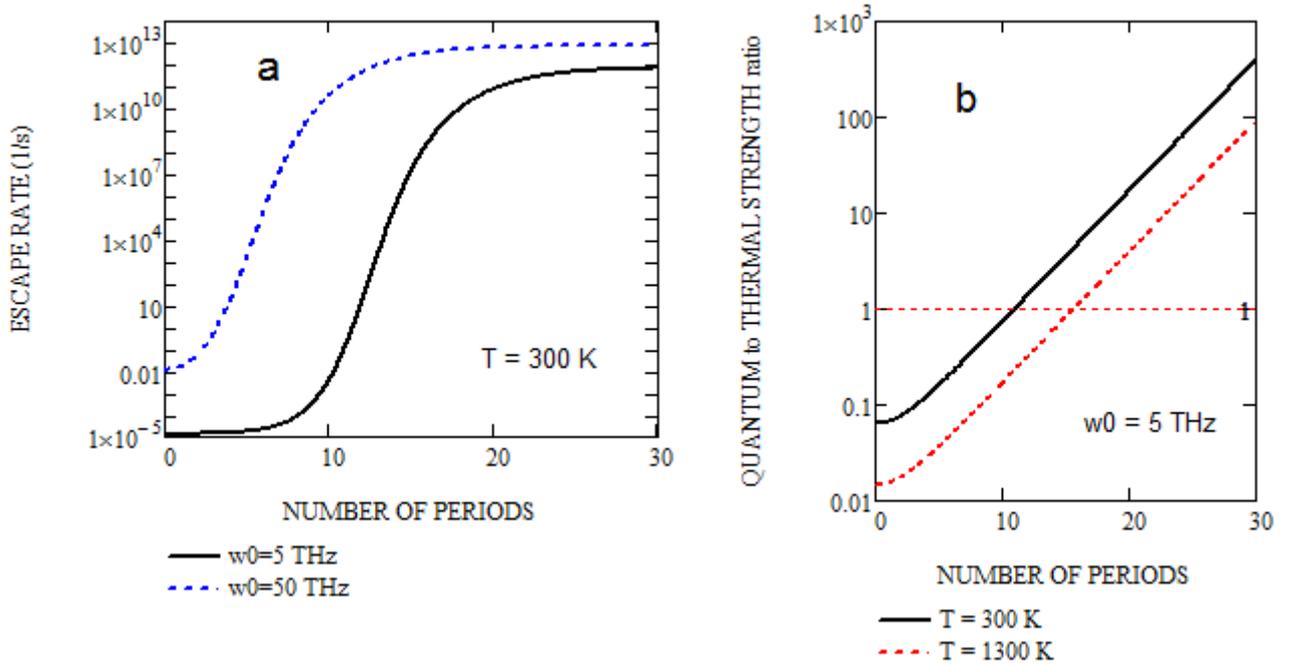

Fig. 4. (a) Escape rate from a well of a depth $E_0 = 1$ eV, eigenfrequency 5 THz (solid curve) and 50 THz (dashed curve) at 300 K vs. the number of oscillation periods in the parametric regime $\Omega = 2\omega_0$ at $g = 0.1$. (b) Increase of the quantum to thermal noise strength, $E_{ZPO}/k_bT$, with increasing number of oscillation periods at different temperatures 300 K and 1300 K.

Such a well depth is typical for many chemical reactions, and the lower frequency limit of ~5 THz is typical for the oscillation frequencies of metal (heavy) atoms while 50 THz is more close to the frequencies of light atoms such hydrogen etc. imbedded in a crystal lattice of more heavy atoms. A characteristic example is hydrogen or deuterium atoms in metal hydrides/deuterides, such as NiH or PdD, in the vicinity of *gap breathers* - a subclass of LAV arising in a regular lattice [15]. The large mass difference between H or D and the metal atoms provides a gap in phonon spectrum, in which gap breathers can be excited either by thermal fluctuations at elevated temperatures or by external driving such as irradiation at low temperatures [16]. As has been argued in [15], the interplay between harmonic and anharmonic forces operating in the gap breather can result in a parametric driving of the potential wells of neighboring light atoms with a double frequency in relation to their eigenfrequencies. Accordingly, various reactions involving hydrogen atoms can be greatly accelerated by external energy input producing LAVs, in agreement with experimental results [17].

Another important consequence of the parametric driving of the well eigenfrequency is the increase of the quantum noise strength as compared to the thermal noise strength represented by the ratio $E_{ZPO}/k_bT$ that increases with increasing number of oscillation periods (Fig. 4b). It means that one can expect quantum effects to dominate over the thermal ones even at elevated temperatures, which may be manifested by a strong deviation from the Arrhenius law.

The parametric driving $\Omega = 2\omega_0$ considered above requires rather special conditions similar to those in gap breathers in diatomic crystals [15], while in many other systems [7], e.g. in metals [8-10], oscillations of atoms in a discrete breather have different amplitudes but the same frequency. This case is more close to the driving of the potential well positions with $\Omega = \omega_0$, which also results in increasing mean energy of the quantum oscillator (eq. (32)), but it does not increase the quantum noise strength since the wave packet dispersion remains constant (eq. (33)). Accordingly, one could expect an acceleration of the escape from a well to occur due to the effective decrease of the well depth given by

$$R_K = \frac{\omega_0}{2\pi}\exp\left[-\left(E_0 - \langle E \rangle\right)/D(T)\right], \quad D(T) = \frac{\hbar\omega_0}{2}\coth\left(\hbar\omega_0/2k_B T\right), \tag{34}$$

where $\langle E \rangle$ is the mean oscillator energy increasing with time according to eq. (32). The corresponding rate of escape from a well with a given parameters in this driving regime is presented in Fig. 5 for a potential well with a depth of 1 eV. It can be seen that the oscillator energy gradually increases up to the activation energy value resulting in a significant increase of the reaction rate, especially at low temperatures.

Note that in a real situation, anharmonicity of the potential well could limit the energy gained by the driving, which should be taken into account in order to make quantitative evaluations of the reaction acceleration by the present mechanism.

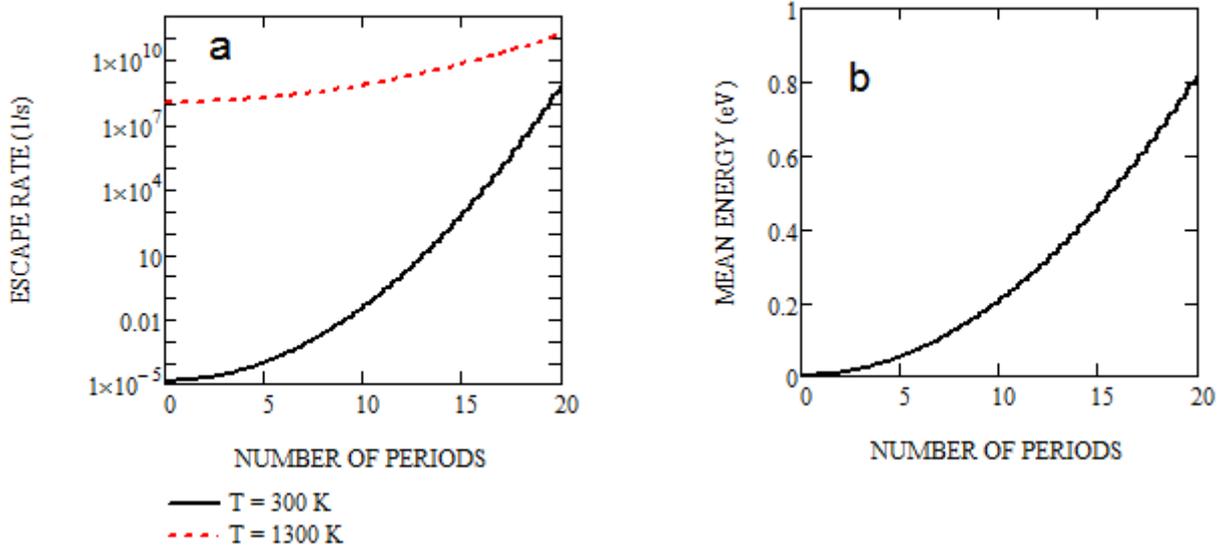

Fig. 5. Escape rate from a well of a depth $E_0 = 1$ eV, eigenfrequency 5 THz at different temperatures (a) and the oscillator mean energy (b) vs. the number of oscillation periods at $\Omega = \omega_0$, $g_A = 0.5$.

## 5. Conclusions and outlook

- Analytical solution of the Schrödinger equation for a periodically driven harmonic oscillator is derived.

- The oscillator zero-point energy is shown to increase in response to parametric modulation of the oscillator eigenfrequency at $\Omega = 2\omega_0$. Based on that, a drastic increase of the escape rate with increasing number of modulation periods is demonstrated in the framework of the modified Kramers theory, which takes into account the quantum noise strength that increases due to the time-periodic driving.

- Time-periodic driving of the potential well positions at $\Omega = \omega_0$ results in increasing mean energy of the quantum oscillator at a constant dispersion of the wave packet. It results in lowering of the effective activation barrier, which may amplify the escape rate significantly, especially at low temperatures.

- The analysed driving modes can be induced by LAVs that can be excited in a crystal bulk or at crystal defects either thermally or by external triggering, which can result in strong catalytic effects. Further investigations in this field based on atomistic modeling of LAV excitation in solids may open the ways of *engineering* of new catalysts.